\documentclass[12pt]{article}
\usepackage{epsfig}

\usepackage{amssymb}
\usepackage{amsmath}
\usepackage{amsfonts}
\usepackage{amsthm}

\theoremstyle{theorem}

\theoremstyle{definition}

\def\bp{\begin{proof}}
\def\ep{\end{proof}}

\def\be{\begin{equation}}
\def\ee{\end{equation}}
\def\ba{\begin{array}{c}}
\def\ea{\end{array}}

\def\ben{$$}
\def\een{$$}

\newcommand{\bea}{\begin{eqnarray}}
\newcommand{\eea}{\end{eqnarray}}

\begin{document}

\titlepage

\vspace{.35cm}

 \begin{center}{\Large \bf

The horizons of observability in ${\cal PT}-$symmetric four-site
quantum lattices

  }\end{center}

\vspace{10mm}

 \begin{center}

 {\bf Miloslav Znojil}

 \vspace{3mm}
Nuclear Physics Institute ASCR,

250 68 \v{R}e\v{z}, Czech Republic

{e-mail: znojil@ujf.cas.cz}

\vspace{3mm}


\end{center}

\vspace{5mm}


\section*{Abstract}

One of the key merits of ${\cal PT}-$symmetric (i.e., parity times
time reversal symmetric) quantum Hamiltonians $H$ lies in the
existence of a horizon of the stability of the system.
Mathematically speaking this horizon is formed by the boundary of
the domain ${\cal D}^{(H)} \subset \mathbb{R}^D$ of the (real)
coupling strengths for which the spectrum of energies is real and
non-degenerate, i.e., in principle, observable. It is shown here
that even in the elementary circular four-site quantum lattices with
$D=2$ or $D=3$ the domain of the hidden Hermiticity  ${\cal
D}^{(H)}$ proves multiply connected, i.e., topologically nontrivial.

\newpage

\section{Introduction}

One of the most interesting formulations of the standard and robust
dictum of quantum mechanics emerged in connection with the
acceptance of the so called ${\cal PT}-$symmetric operators of
observables where ${\cal P}$ means parity while ${\cal T}$
represents time reversal (cf. review papers
\cite{Carl,Doreyali,SIGMA} for an exhaustive discussion). One of the
main reasons of the last-year rebirth of interest in this new
paradigm may be seen, paradoxically, in its impact on classical
experimental optics \cite{Makris}.

The latter experimental activities (i.e., basically, the emergence
of a few successful classical-physics simulations of quantum
effects) re-attracted attention to the innovative theory. We may
mention, {\it pars pro toto}, paper \cite{Wang} which offered an
{\em exhaustive} constructive classification of {\em all} of the
${\cal PT}-$symmetric quantum Hamiltonians $H$ defined in the {\em
finite-dimensional} Hilbert spaces ${\cal H}$ of dimensions $N=2$
and $N=3$.

Inside the most elementary $N \leq 3$ family of models no real
surprises and spectral irregularities have been encountered. In
contrast, in Ref.~\cite{anom} we found that certain anomalies
certainly emerge at $N=8$. In our present brief continuation of
these developments we intend to show that the simplest models
exhibiting similar irregularities in their spectra already occur,
unexpectedly, as early as at the next Hilbert-space dimension $N=4$.

\section{The four-site quantum-lattice models}

\begin{figure}[htb]                     
\begin{center}                         
\epsfig{file=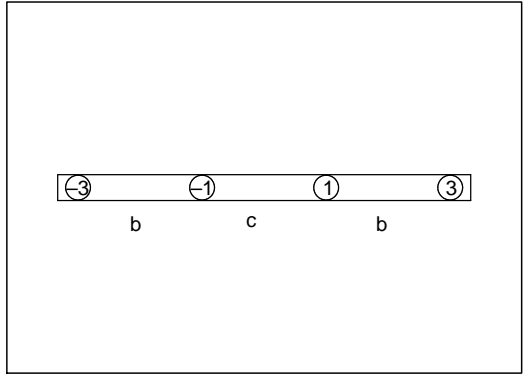,angle=270,width=0.5\textwidth}
\end{center}                         
\vspace{-2mm}\caption{Graphical symbol for the straight-line
open-end four-site lattice. The numbers in the small circles (=
sites) are the unperturbed energies while the letters $b$ and $c$
near the nearest-neighbor-interaction lines represent the (real)
couplings.
 \label{fi1}}
\end{figure}

\subsection{The exactly solvable straight-line case.}

In Refs.~\cite{horizon} the successful tractability of
more-than-three-dimensional Hamiltonian matrices resulted from a
drastic simplification of their structure. We merely admitted their
tridiagonal versions. In the language of physics this corresponded
to the picture in which the system lived on a  $N-$site
straight-line lattice endowed with the mere nearest-neighbor
interactions. At $N=4$ this is schematically depicted in
Fig.~\ref{fi1}. The small circles represent the sites while their
frame-line connections symbolize the interactions.

The left-right symmetric straight-line lattice of Fig.~\ref{fi1}
(i.e., of Refs.~\cite{horizon}) is being assigned the Hamiltonian
given in the form of the two-parametric real matrix
 \be
 H=H^{(4)}(b,c)=\left[ \begin {array}{cccc}
  -3&b&0&0
 \\\noalign{\medskip}-b&-1&c&0
 \\\noalign{\medskip}0&-c&1&b
 \\\noalign{\medskip}0&0&-b&3\end {array}
 \right]\,.
 \ee
The quantitative analysis of the models of this form is more or less
trivial even at larger $N>4$. Curious reader may find many details,
say, in review paper \cite{acta}.

\subsection{${\cal PT}-$symmetric circular lattices and their simplest four-site example}

\begin{figure}[htb]                     
\begin{center}                         
\epsfig{file=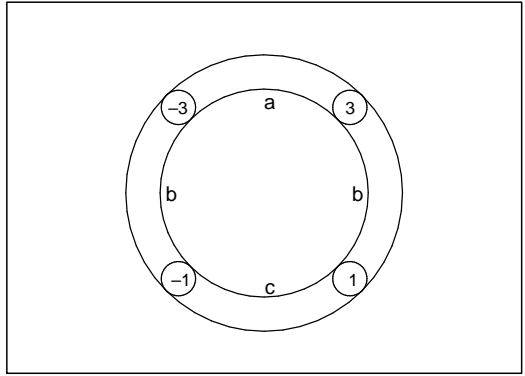,angle=270,width=0.5\textwidth}
\end{center}                         
\vspace{-2mm}\caption{The circular four-site lattice.
 \label{fi3}}
\end{figure}

 \noindent
Once we replace Fig.~\ref{fi1} by its circular version of
Fig.~\ref{fi3} we may immediately interpret the new diagram as
representing the new $N=4$ quantum model which is given by the
following three-parametric four-by-four matrix form of the
Hamiltonian as proposed and studied in  Ref.~\cite{periodic},
 \be
 H=H^{(4)}(a,b,c)=\left[ \begin {array}{cccc} -3&b&0&-a
 \\\noalign{\medskip}-b&-1&c&0
 \\\noalign{\medskip}0&-c&1&b
 \\\noalign{\medskip}a&0&-b&3\end {array}
 \right]\,.
 \ee
It may be shown that in the new model with one more coupling which
connects the ``upper two" sites, the method of construction of the
boundary $\partial {\cal D}^{(H)}$ does not change. At any number
$N$ of sites along the (circular) lattice the reality property of
the spectrum of the energies will  remain tractable by the standard
mathematical techniques, a few representative samples of which may
be found in Ref.~\cite{nje11}. Interested readers may also search
for a broader mathematical context, say, in Refs.~\cite{ams,amsdve}.

Once we restrict our attention just to our special toy model
$H^{(4)}(a,b,c)$ it proves sufficient to recall the entirely
elementary considerations of Ref.~\cite{nje11}. This leads to the
conclusion that the spectrum of energies stays real and
nondegenerate if and only if the triplet of parameters $(a,b, c)$
lies inside the domain
 \be
 {\cal D}^{(H)}:=\left \{  (a,b,c) \in \mathbb{R}^3\,\left
 |\,W(a,b,c) > 0\,,\ Q(a,b,c) > 0\,,\ P(a,b,c) > 0\
 \right . \right \}\,
 \ee
where
 \be
 W(a,b,c) 
 =\left (8+c^2-a^2 \right )^2-4\,\left [16-(a+c)^2 \right ]\,b^2\,,
 \label{eq4}
 \ee
 \be
 Q(a,b,c)
 =\left[ (a+3)(c-1)-{b}^{2} \right]  \left[ (a-3)(c+1)-{b}^{2}
 \right]\,
 \ee
and
 \be
 P(a,b,c)=10-{a}^{2}-2\,{b}^{2}-{c}^{2}\,.
 \ee
In other words, for the couplings moving to the two-dimensional
surfaces of ${\cal D}^{(H)}$ from inside one observes that the
quadruplets of the real bound-state energies themselves behave in an
easily understandable manner. The reason is that we may rewrite the
secular equation in the form ${\cal S}(s,a,b,c)=0$ where the
energies $E_\pm=\pm \sqrt{s}$ emerge in pairs and where
 $$
 {\cal S}(s,a,b,c):=s^{2}+ \left( -10+{c}^{2}+2\,{b}^{2}
 +{a}^{2} \right) \,s +9+6\,{b}^{2}-9\,{c}^{2}+{b}^{4}-2\,ca{b}^{2}
 -{a}^{2}+{c}^{2}{a}^{2}\,.
 $$
This recipe generates the two auxiliary roots
 $$
 4\,s=4\,s^{(\pm)}={20-2\,{a}^{2}-2\,{c}^{2}-4\,{b}^{2}\pm 2\,\sqrt
 {W(a,b,c)}}
 $$
where we already know the function of Eq.~(\ref{eq4}),
 $$
 W(a,b,c)=64+16\,{c}^{2}-64\,{b}^{2}-16\,{a}^{2}
 +{c}^{4}+4\,{c}^{2}{b}^{2}-2\,{c}^{2}{a}^{2}
 +4\,{b}^{2}{a}^{2}+{a}^{4}+8\,ca{b}^{2}\,.
 $$
In the spirit of the general results of Ref.~\cite{periodic} we may
summarize that

\begin{enumerate}

\item
whenever $W(a,b,c)\to 0^+$ one can spot the two pairs of energies
which approach the two distinct values
 \ben
  E^{(W=0)}_\pm =\pm \sqrt{(10-{a}^{2}-2\,{b}^{2}-{c}^{2})/2}\,
 \een
representing the two limiting  doubly-degenerate energies;

\item
whenever $Q(a,b,c)\to 0^+$ just the two energies move to zero while
the other two energies do not vanish in general,
 \ben
  E_{0,3} =\pm \sqrt{10-{a}^{2}-2\,{b}^{2}-{c}^{2}}\,;
 \een

\item
for $P(a,b,c)\to 0^+$ we must expect that all of the four real
energies will vanish simultaneously.

\end{enumerate}

\section{Two-parametric simplified versions of the circular
four-site lattice}

\subsection{The case of $a=0$}

\begin{figure}[htb]                     
\begin{center}                         
\epsfig{file=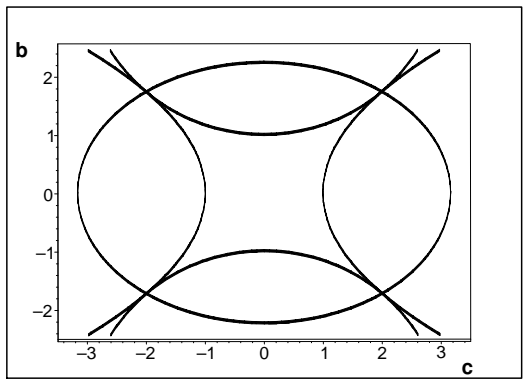,angle=270,width=0.5\textwidth}
\end{center}                         
\vspace{-2mm}\caption{The graphical determination of the innermost
star-shaped domain ${\cal D}^{(H)}$ as assigned to the quantum
lattice of Fig.~\ref{fi1} in Refs.~\cite{horizon}.
 \label{fi2}}
\end{figure}
 \noindent
Naturally,  in the no-upper-interaction limit $\lim_{a \to
0}\,H^{(4)}(a,b,c)=H^{(4)}(b,c)$ we return to the elementary
straight-line model of Fig.~\ref{fi1}. For our present purposes it
is then sufficient to recall that the main features of such a
simplified model were described in Refs.~\cite{horizon}. In
particular, we know that at $N=4$ the spectrum of energies remains
real and nondegenerate inside the innermost star-shaped domain
${\cal D}^{(H)} \subset \mathbb{R}^2$ shown, in Fig.~\ref{fi2}, as
lying inside an auxiliary circumscribed ellipse. The boundary
$\partial {\cal D}^{(H)}$ (i.e., the physical horizon of the system
in question) is composed of the four hyperbola-shaped curves. The
key features of this example (like the triple intersections of the
boundaries, etc.) generalize, {\em mutatis mutandis}, to the family
of the similar models at all of the dimensions
$N<\infty$~\cite{acta}.

We are now prepared to replace the elementary and transparent
graphical determination of the star-shaped domain ${\cal D}^{(H)}$
assigned to the straight-line quantum lattice and displayed in
Fig.~\ref{fi2} by its much more complicated $a \neq 0$ analogue. At
a freely variable $a$  the knowledge of the $a=0$ section may and
will still serve us as a very useful independent test of our
forthcoming observations and conclusions.

\subsection{The case of $b=0$}

\begin{figure}[htb]                     
\begin{center}                         
\epsfig{file=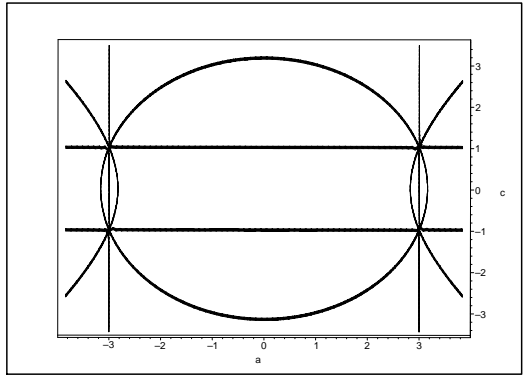,angle=270,width=0.5\textwidth}
\end{center}                         
\vspace{-2mm}\caption{The degenerate case of the simply connected
{\em rectangular} domain ${\cal D}^{(H)}$ at $b=0$.
 \label{fi5}}
\end{figure}

 \noindent
In another preparatory step we notice that our present four-site toy
model $H^{(4)}(a,b,c)$ degenerates to the trivial non-interacting
composition (i.e., direct sum) of the two $N=2$ models at $b=0$. For
this reason the $b=0$ limiting case should be considered
exceptional.

It is worth emphasizing that at the $b=0$ two-dimensional special
case, even the general definition of the domain ${\cal D}^{(H)}$ is
slightly misleading. Indeed, Fig.~\ref{fi5} which displays the three
sets of boundaries (viz., the two hyperbolas $W(a,0,c)=0$, the four
straight lines $Q(a,0,c)=0$ and the single circle $P(a,0,c)=0$,
respectively) should not be taken too literally because one of the
boundaries (viz.,  the doublet of hyperbolas $W(a,0,c)=0$) describes
in fact a sign-non-changing (i.e., the reality-of-energies
non-changing) curve of the doubly degenerate (and, hence, irrelevant
and removable) zeros of the function $W(a,0,c)=(8+c^2-a^2)^2$.

This means that at $b=0$ the domain ${\cal D}^{(H)}$ is strictly
rectangular and strictly simply connected. In this context one of
the key messages of our present study is the surprising discovery of
the loss of {\em both} of these properties in the general case with
the freely variable coupling strength $b$.

Incidentally, the multinomial $W(a,b_{spec},c)$ becomes factorizable
also at $b_{spec}=1$,
 $$
 W(a,1,c)=
\left( a+c \right)  \left(
{a}^{3}-c{a}^{2}-12\,a-{c}^{2}a+20\,c+{c}^{3} \right)\,.
 $$
This is an artifact which does not carry any immediate physical
meaning. Its manifestation is of a purely geometrical character
which will only be briefly mentioned later.

\begin{figure}[htb]                     
\begin{center}                         
\epsfig{file=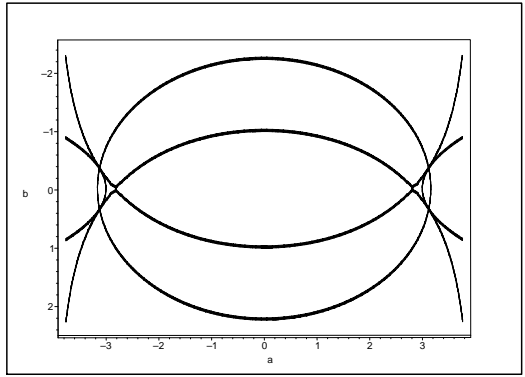,angle=270,width=0.5\textwidth}
\end{center}                         
\vspace{-2mm}\caption{The triply connected nature of the
 triple-overlap domain ${\cal D}^{(H)}$ for the
quantum lattice of Fig.~\ref{fi3} at
 $c=0$ (i.e., in the
no-central-coupling extreme).
  \label{fi4}}
\end{figure}

\subsection{The case of $c=0$ \label{juraj}}

\begin{figure}[htb]                     
\begin{center}                         
\epsfig{file=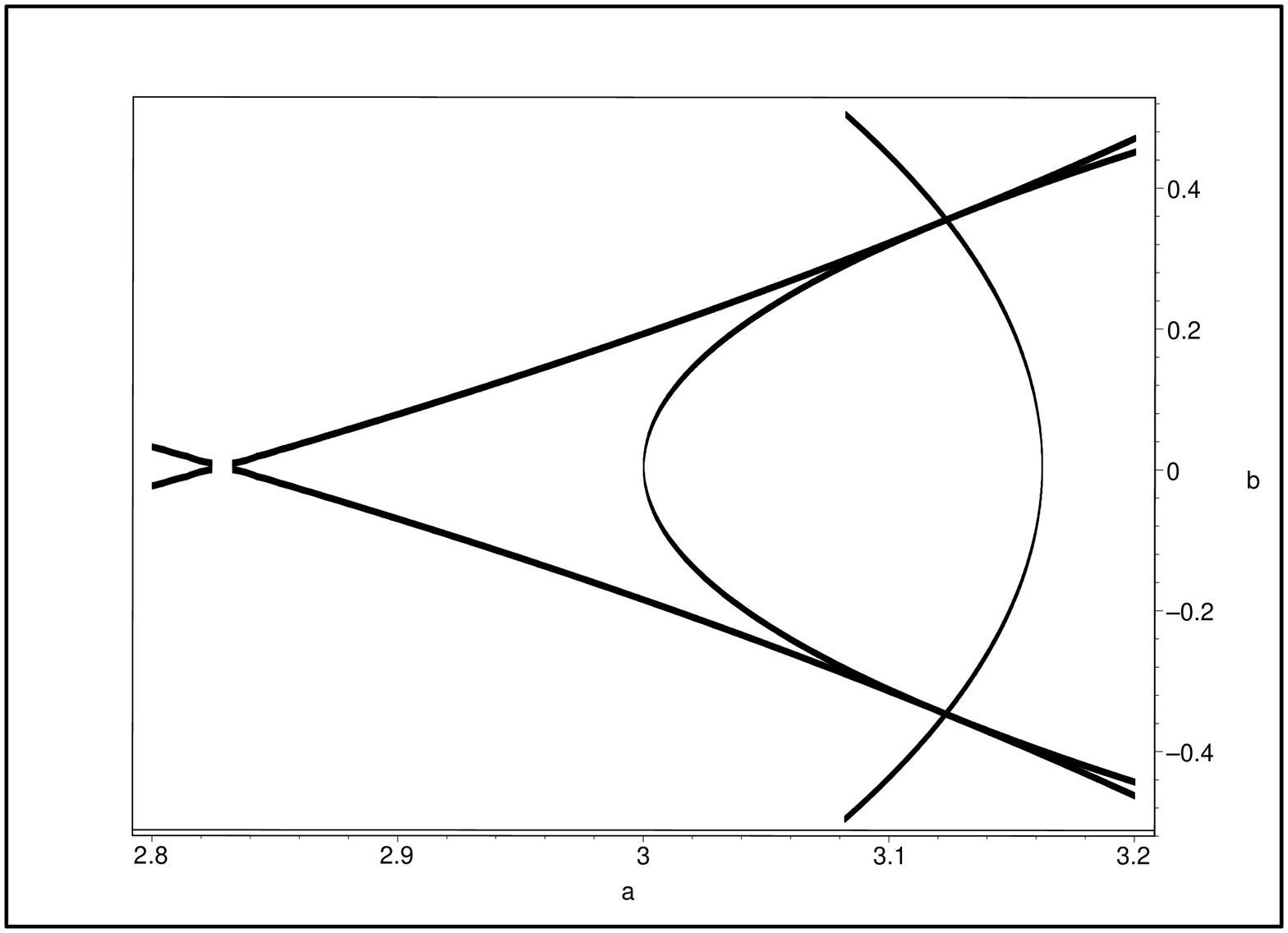,angle=270,width=0.5\textwidth}
\end{center}                         
\vspace{-2mm}\caption{Same as Fig.~\ref{fi4} (detail).
  \label{fi4a}}
\end{figure}

\begin{figure}[htb]                     
\begin{center}                         
\epsfig{file=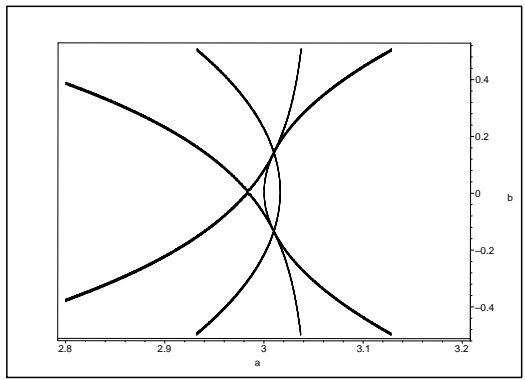,angle=270,width=0.5\textwidth}
\end{center}                         
\vspace{-2mm}\caption{The deformation of Fig.~\ref{fi4a} at
$c=0.95$.
  \label{fi4aa}}
\end{figure}

 \noindent
In the third (and last) preparatory step let us discuss the
vanishing-coupling special case in which $c = 0$ and
 \be
 W(a,b,0)
 =\left (8-a^2 \right )^2-4\,\left [16-a^2 \right ]\,b^2\,,
 \label{oeq4}
 \ee
 \be
 Q(a,b,0)
 =\left(3+{b}^{2}+a \right)  \left( 3+{b}^{2}-a
 \right)\,
 \ee
and
 \be
 P(a,b,0)=10-{a}^{2}-2\,{b}^{2}\,.
 \ee
The detailed study of precisely this special case reveals in fact
the possibility of the emergence of a topological nontriviality in
the general case. The detailed form of such a $c = 0$ hint may be
seen in Fig.~\ref{fi4} and in its magnified version~\ref{fi4a}.
Indeed, as long as the condition $Q(a,b,0)>0$ degenerates to the
elementary constraint
 \ben
  3+{b}^{2} > a >
 -  3-{b}^{2}
 \,
 \een
just the left and right small horizontal-parabolic segments (with
their extreme at $b=0$ and $|a_{max}|=3$) should be cut out of the
elliptic domain with $P(a,b,0)>0$ as inadmissible since $Q(a,b,0)<0$
there. As long as we only have $16>a^2$, the remaining constraint
$W(a,b,0)>0$ acquires the elementary form
 \be
 |b|<\frac{1}{4}\frac{\left |8-a^2 \right |}{\sqrt{16-a^2}}\,
 \ee
of the geometric limitation of the admissible range of $b$ by the
two intersecting or rather broken and touching curves. Indeed, one
must keep in mind that the nonnegative function $W(a,0,0)=(8-a^2)^2$
solely vanishes at $a^2=8$.

These observations imply that the allowed region decays into the
three disconnected open sets (cf. Figs.~\ref{fi4} -- \ref{fi4aa}).
The big one is formed by the eye-shaped vicinity of the origin, with
its extremes at the points $(a,b)_\pm =(\pm \sqrt{8},0)$. The other
two smaller open sets are fish-tail-shaped. In the pictures (cf., in
particular, the magnified right one in Fig.~\ref{fi4a}) these two
domains are easily spotted as containing the respective $b=0$
intervals of $|a| \in (\sqrt{8},3)$. In this sense they may be
expected to support a perturbatively inaccessible ``strong-coupling"
dynamical regime.

An additional indication of the suspected emergence of topological
as well as dynamical nontrivialities is offered by Fig.~\ref{fi4aa}
where the same separation of a strong-coupling piece of the domain
${\cal D}^{(H)}$ is shown to survive up to the very extreme of
$c\approx 1$.


\section{The domain of cryptohermiticity in the full-fledged
three-parametric dynamical scenario}

\begin{figure}[htb]                     
\begin{center}                         
\epsfig{file=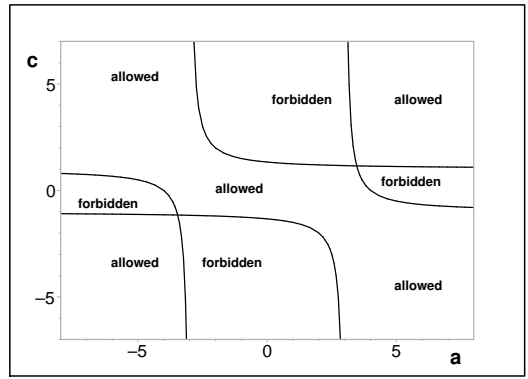,angle=270,width=0.5\textwidth}
\end{center}                         
\vspace{-2mm}\caption{Boundaries $Q(a,b,c)=0$ and forbidden parts of
the $a-c$ plane as sampled at $b^2=1$.
  \label{faa}}
\end{figure}

\subsection{The auxiliary domains and their boundaries}


 \noindent
The domain  ${\cal D}^{(H)}$ of parameters for which our Hamiltonian
$H^{(4)}(a,b,c)$ generates the unitary evolution of quantum system
is defined as an intersection of the triplet of domains ${\cal
D}^{(P,Q,W)}$ in $\mathbb{R}^3$. After the preparatory
considerations as presented in the preceding sections let us now
leave all of the three parameters $ a$, $b$ and $ c$ freely variable
and recall that

\begin{itemize}

\item
the domain ${\cal D}^{(P)}$ is defined by the inequality
$P(a,b,c)=10-{a}^{2}-2\,{b}^{2}-{c}^{2}>0$. It is compact so that we
may restrict our attention just to the intervals of $b^2<5$,
$a^2<10$ and $c^2<10$. At any fixed $b^2<5$ the section of this
first auxiliary domain coincides with the interior of a central
circle in $a-c$ plane with radius $R = \sqrt{10-2b^2}$;

\item
the allowed interior of the triply connected domain ${\cal D}^{(Q)}$
is defined by the inequality $Q(a,b,c) =\left[ (a+3)(c-1)-{b}^{2}
\right]  \left[ (a-3)(c+1)-{b}^{2} \right]$. In the $a-c$ plane the
boundaries of this domain are two hyperbolas sampled at $b^2=1$ in
Fig.\ref{faa};

\item
the third auxiliary domain ${\cal D}^{(W)}$ is defined by the
inequality $W(a,b,c) =\left (8+c^2-a^2 \right )^2-4\,\left
[16-(a+c)^2 \right ]\,b^2>0$.

\end{itemize}

 \noindent
The description of the latter domain ${\cal D}^{(W)}$ is slightly
less trivial. It may be based on the observation that the interior
of this domain covers all the exterior of the strip where $|a+c|>4$.
Then the interior of this strip may be reparametrized,
 \ben
  c-a=2\,\tau(c,a) \in (-\infty,\infty)\,,\ \ \ \ \
  c+a=4\,\sin \varphi(c,a)\,, \ \ \ \varphi(c,a) \in (-\pi/2,\pi/2)\,
 \een
making the rest of the domain ${\cal D}^{(W)}$ determined by the
elementary inequality
 \be
 |b| <
 \frac{|1+\tau(c,a) \sin [\varphi(c,a)]|}{\cos[ \varphi(c,a)]}
 \,.
 \ee
This means that within the restricted range of $\tau(c,a) \in
(-\sqrt{10},\sqrt{10})$ the growth of $|b|\to \infty$ must be
compensated by the decrease of $\cos[ \varphi(c,a)]\to 0$, i.e., by
the convergence $c \to \pm 1-a$. This makes the strip-restricted
part of the domain ${\cal D}^{(W)}$ very small but increasing with
the decrease of $|b|$ from a sufficiently large initial value.

\subsection{The boundaries of the cryptohermiticity domain}

In the light of the latter comment it makes good sense to start the
study of the ``physical" overlaps ${\cal D}^{(W)}$ of the three
auxiliary domains ${\cal D}^{(P,Q,W)}$ at the maximal admissible
plane of $b=b^{(P)}=\sqrt{5}$ which touches the boundary $\partial
{\cal D}^{(P)}$ at $a=c=0$. This points still lies out of the
domains ${\cal D}^{(W)}$ and ${\cal D}^{(H)}$ since
$W(0,\sqrt{5},0)=64-320<0$. In a search for the first touch between
the $b-$plane and boundary $\partial {\cal D}^{(H)}$ we must
diminish our $b$ and move into the interior of ${\cal D}^{(P)}$.

\begin{figure}[htb]                     
\begin{center}                         
\epsfig{file=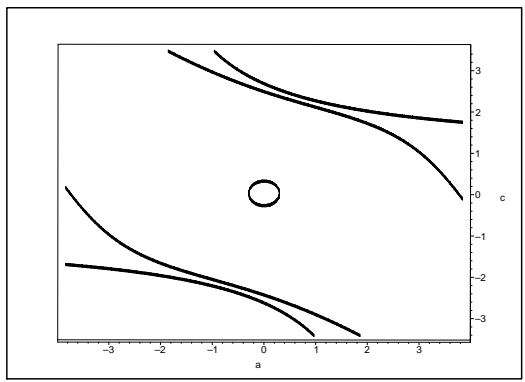,angle=270,width=0.5\textwidth}
\end{center}                         
\vspace{-2mm}\caption{$b=\sqrt{5}-1/100=2.226067977$.
 \label{fig9}}
\end{figure}

In the first illustrative example at $b=\sqrt{5}-1/100$ our
Fig.~\ref{fig9} displays the motion of the triplet of boundaries
$\partial {\cal D}^{(P,Q,W)}$ projected into the $a-c$ real plane.
This picture shows that the corresponding section of the first
domain ${\cal D}^{(P)}$ becomes nonempty. Still, it just occupies
the interior of a very small circle $\mathbf{C}(b) =
\partial {\cal D}^{(P)}|_{b=fixed}$ with the center at the origin.

\begin{figure}[htb]                     
\begin{center}                         
\epsfig{file=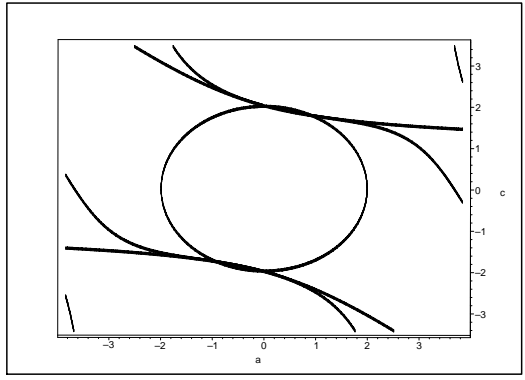,angle=270,width=0.5\textwidth}
\end{center}                         
\vspace{-2mm}\caption{$b=\sqrt{5}-1/2=1.736067977$.
 \label{fig7}}
\end{figure}

The interior of the second domain ${\cal D}^{(Q)}$ is perceivably
bigger since, in the manner indicated by Fig.~\ref{faa} above, it
occupies the large domain between the two outermost, $b-dependent$
hyperbolic curves $\mathbf{H}_{1,2}(b) \subset \partial {\cal
D}^{(Q)}|_{b=fixed}$. The overlap ${\cal D}^{(H)}$ itself remains
empty because the third domain ${\cal D}^{(W)}$ is localized behind
the two remaining and less trivially parametrized curves
$\mathbf{G}_{1,2}(b) \subset
\partial {\cal D}^{(W)}|_{b=fixed}$.

%
%

\begin{figure}[htb]                     
\begin{center}                         
\epsfig{file=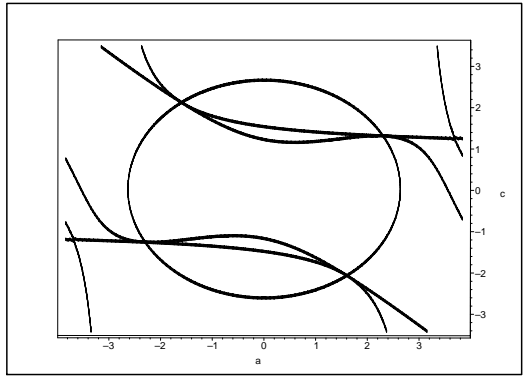,angle=270,width=0.5\textwidth}
\end{center}                         
\vspace{-2mm}\caption{$b=\sqrt{5}-1=1.236067977$.
 \label{fig6}}
\end{figure}

During the subsequent decrease of $b$ sampled by Fig.~\ref{fig9} the
two curves $\mathbf{G}_{j}(b)$ and $\mathbf{H}_{j}(b)$ (assigned the
same subscript $j=1$ or $j=2$) get closer to each other while the
internal circle $\mathbf{C}(b)$ gets larger. At each $j$ and at the
same value of $b$ both the curves $\mathbf{G}_j, \mathbf{H}_j$ touch
the circle $\mathbf{C}(b)$. At a still smaller
$b=\sqrt{5}-1/2=1.736067977$ they already move inside, sharing their
two separate intersections with the circle. This situation is
illustrated in Fig.~\ref{fig7}.

\begin{figure}[htb]                     
\begin{center}                         
\epsfig{file=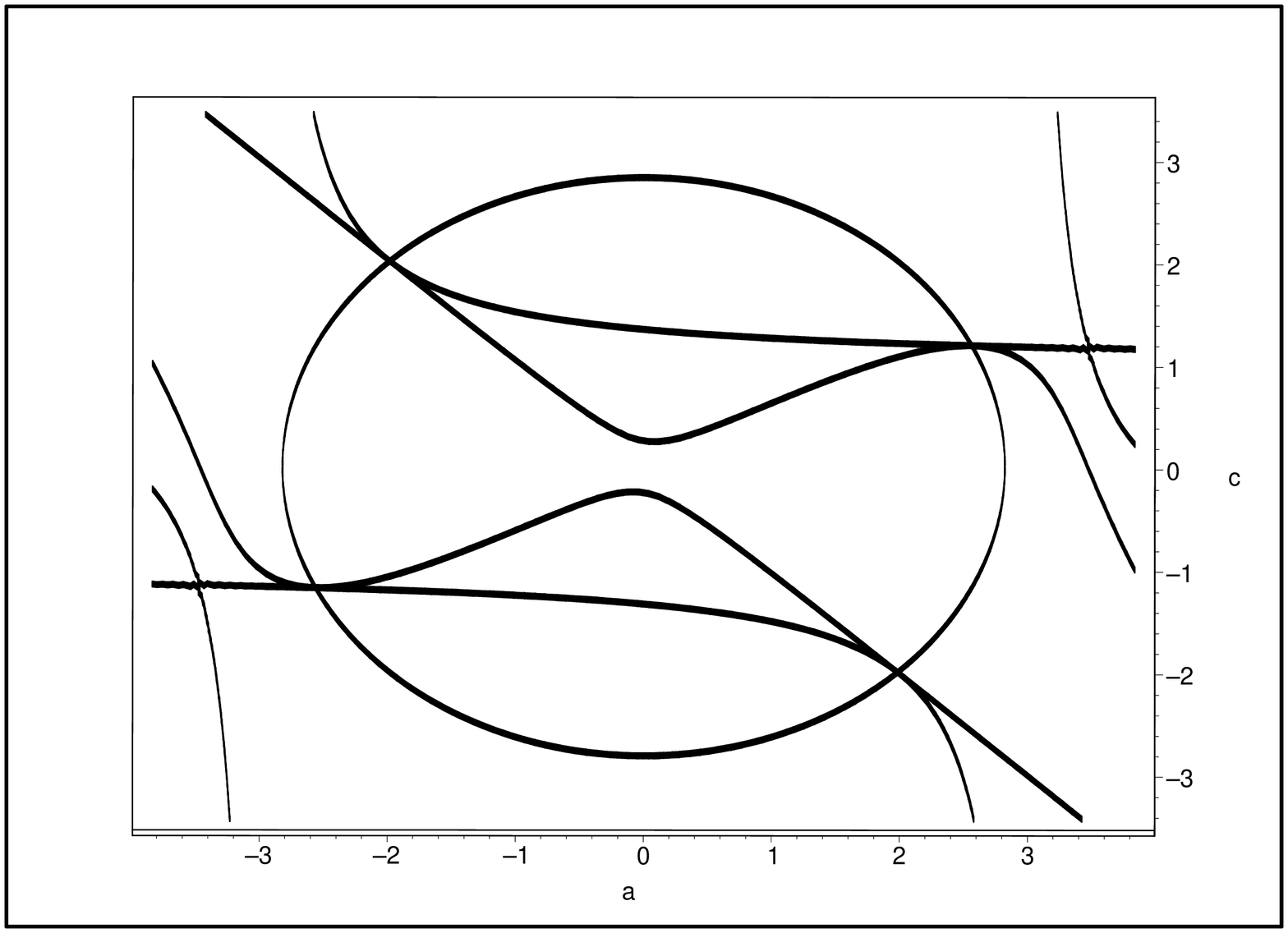,angle=270,width=0.5\textwidth}
\end{center}                         
\vspace{-2mm}\caption{$b=1.01$.
 \label{fig5}}
\end{figure}

\begin{figure}[htb]                     
\begin{center}                         
\epsfig{file=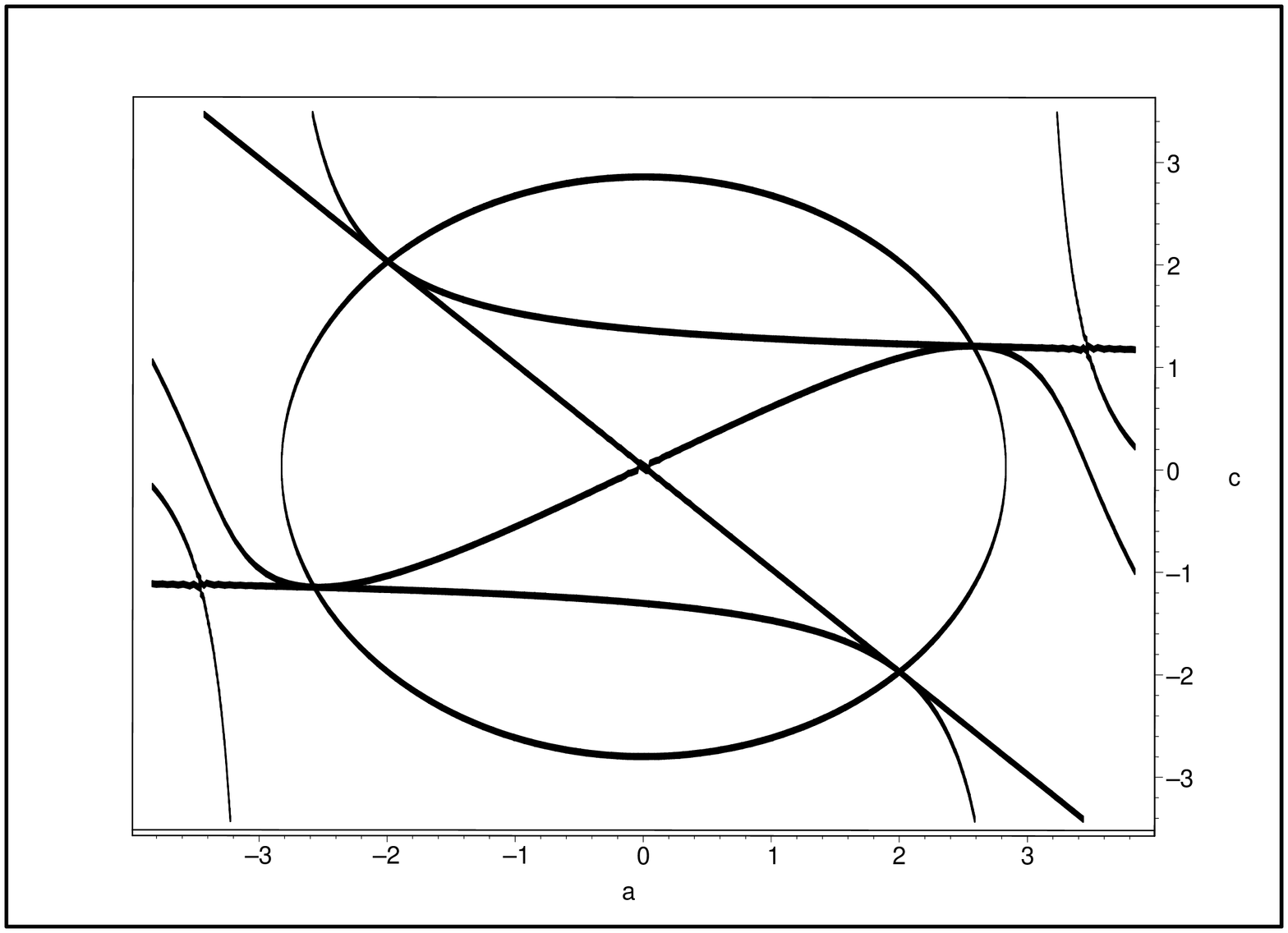,angle=270,width=0.5\textwidth}
\end{center}                         
\vspace{-2mm}\caption{$b=1$.
 \label{fig4}}
\end{figure}

At any $j$ and during the further decrease of the parameter $b$ the
two triple-intersection points between $\mathbf{C}(b)$,
$\mathbf{G}_j$ and $\mathbf{H}_j$ move apart. Between them one
discovers the formation of the first two non-empty components of the
physical domain ${\cal D}^{(H)}$. These components are disconnected,
extremely narrow and eye-shaped, with the ``eyes almost closed" but
``slowly opening" with the further decrease of $b$. Graphically, the
generic situation of this type in illustrated by Figs.~\ref{fig6}
and~\ref{fig5}.

\begin{figure}[htb]                     
\begin{center}                         
\epsfig{file=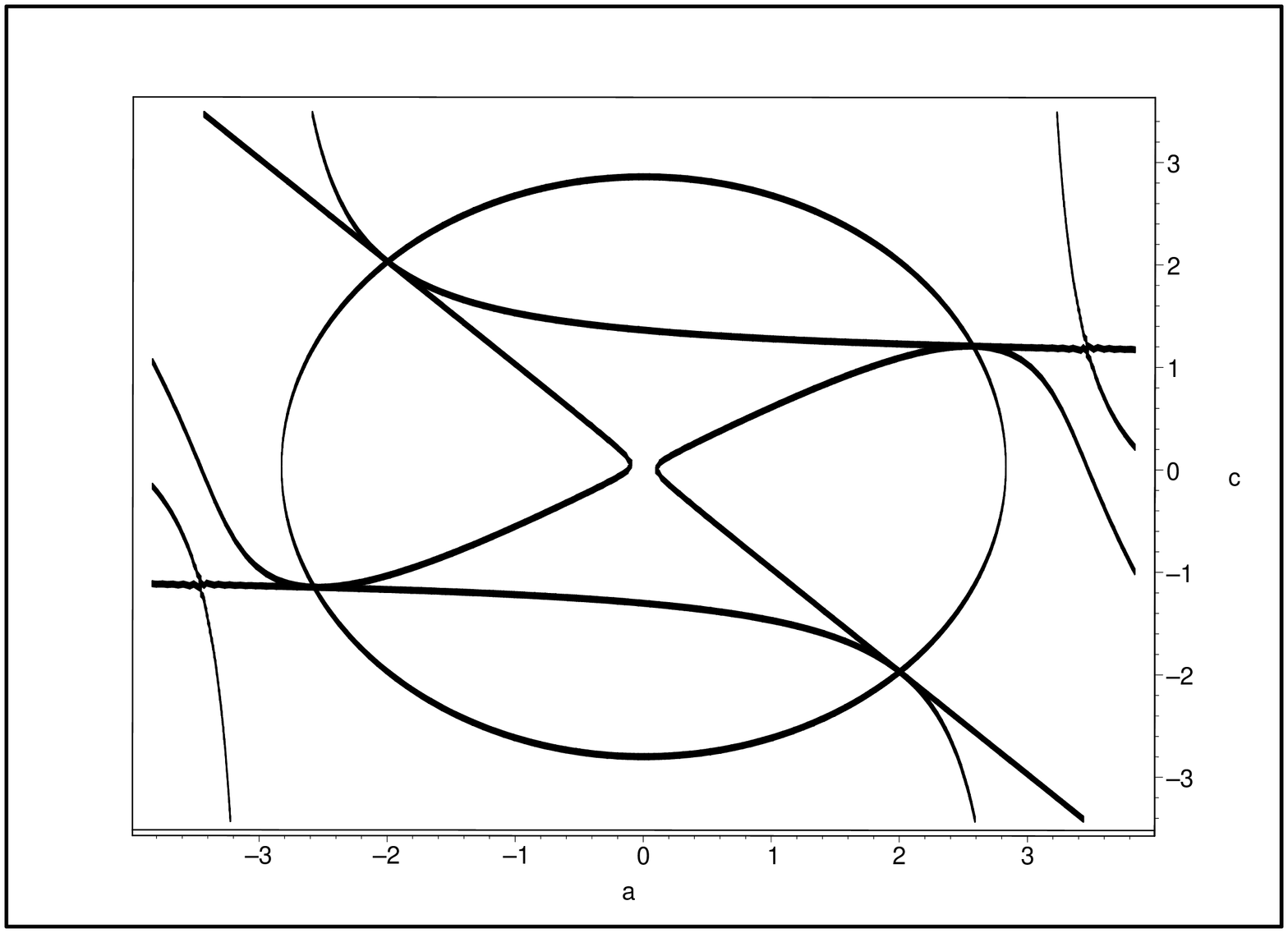,angle=270,width=0.5\textwidth}
\end{center}                         
\vspace{-2mm}\caption{$b=0.999$.
 \label{fig3}}
\end{figure}

The next qualitative change of the pattern occurs at the
above-mentioned special value of $b=1$ as which we touch the saddle
of the surface $\partial {\cal D}^{(W)}$. Slightly before this
happens we encounter the situation depicted in Fig.~\ref{fig5} where
the two separate subdomains of the physical domain ${\cal
D}^{(H)}|_{b=fixed}$ already almost touch. Next they do touch (cf.
Fig.~\ref{fig4}) and, subsequently, get connected (cf. the next
Fig.~\ref{fig3}).

\begin{figure}[htb]                     
\begin{center}                         
\epsfig{file=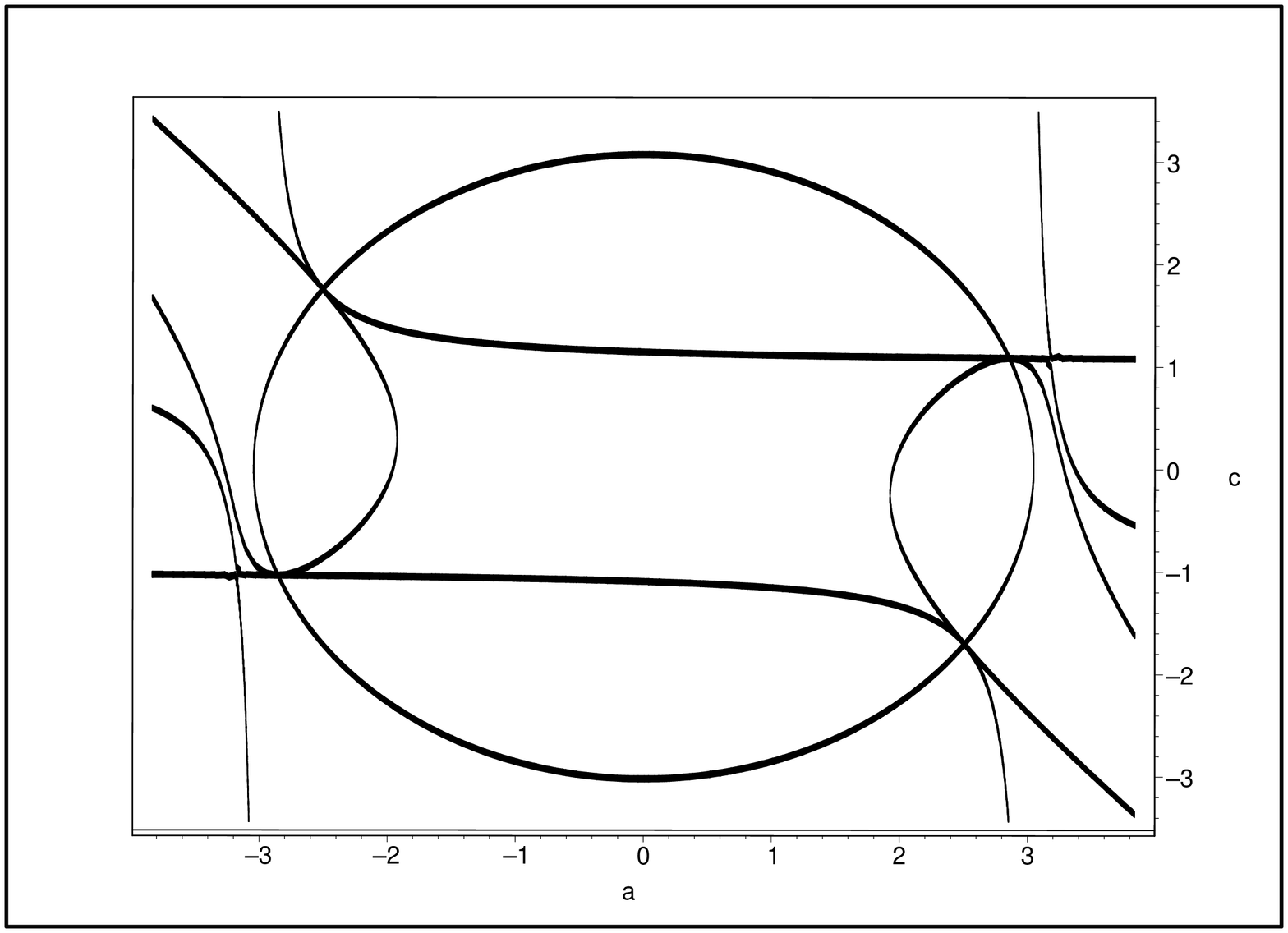,angle=270,width=0.5\textwidth}
\end{center}                         
\vspace{-2mm}\caption{$b=0.6$.
 \label{fi9}}
\end{figure}

Surprisingly enough, below the saddle point $b = 1$ the topological
surprises are still not at the end. There are no real news even at
$b=0.6$ (cf. Fig.~\ref{fi9}). Nevertheless, in the latter picture we
already must pay attention to the two subdomains with the maximal
$a^2$s.

\begin{figure}[htb]                     
\begin{center}                         
\epsfig{file=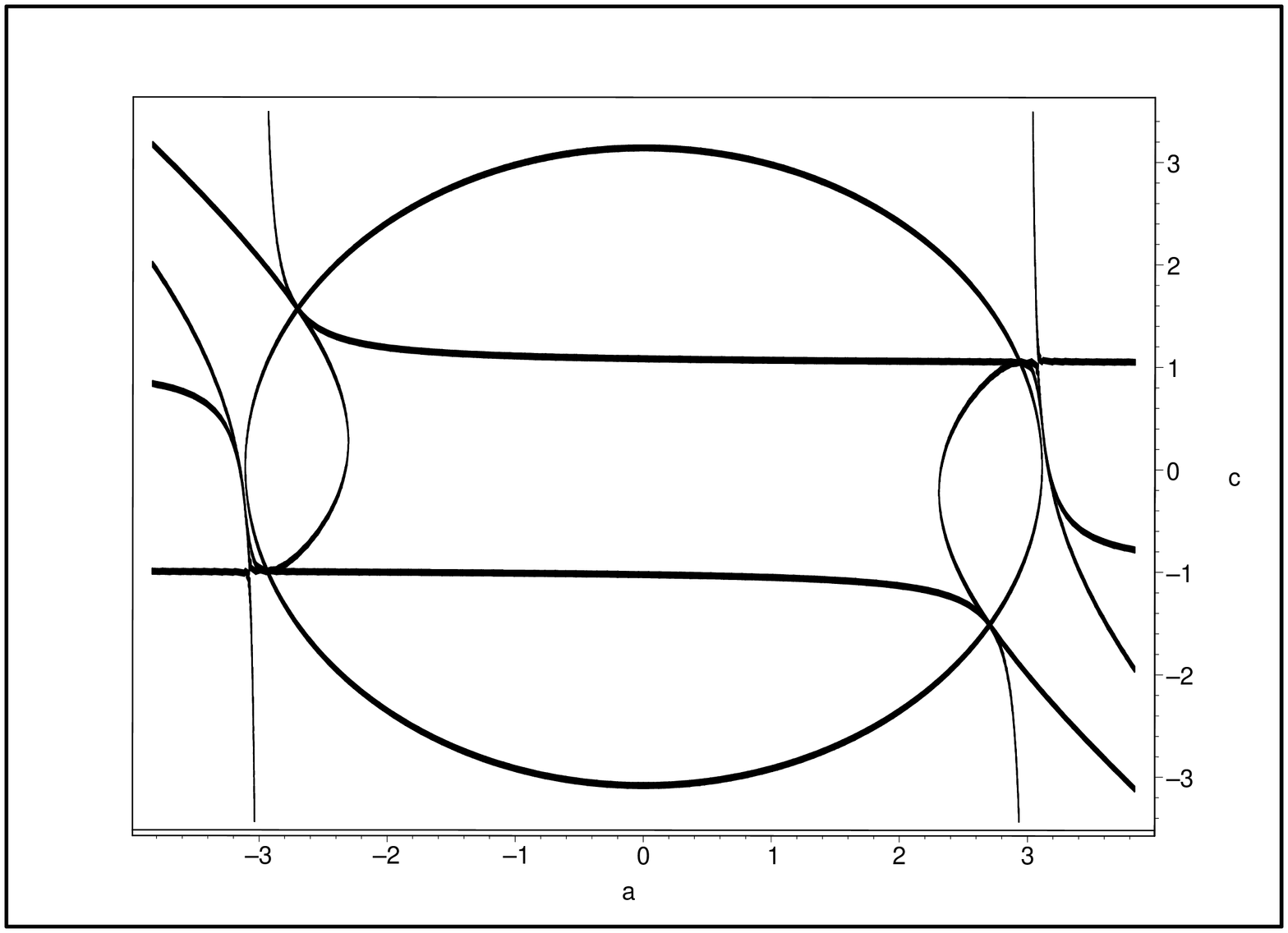,angle=270,width=0.5\textwidth}
\end{center}                         
\vspace{-2mm}\caption{$b=0.4$.
 \label{fi8}}
\end{figure}

Having selected just the right end of the (symmetric) picture at the
positive $a \approx 3$ we reveal the emergence of a tendency towards
a new intersection between the (up to now, safely external and
non-interfering) second branches of the $Q(a,b,c,)-$related
hyperbolas $\mathbf{H}_j^{(second)}$ and of the back-bending
boundaries $\partial {\cal D}^{(W)}|_{b=fixed}$. For example, these
curves get very close to each other but still do not intersect yet
at $b=0.4$, staying also outside of the central circular domain
${\cal D}^{(P)}$ (cf. Fig.~\ref{fi8}).

\begin{figure}[htb]                     
\begin{center}                         
\epsfig{file=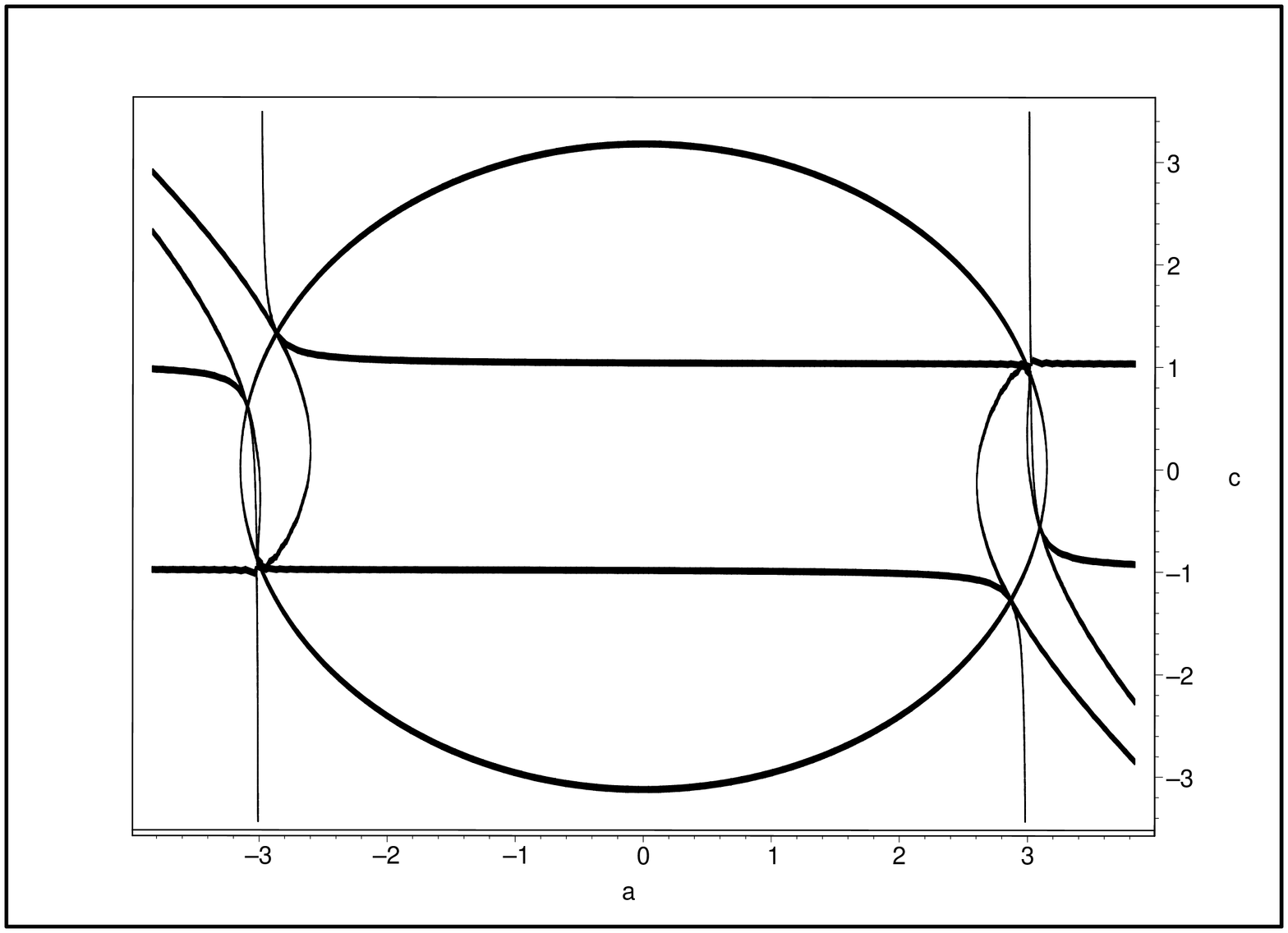,angle=270,width=0.5\textwidth}
\end{center}                         
\vspace{-2mm}\caption{$b=0.2$.
 \label{fi7}}
\end{figure}

\begin{figure}[htb]                     
\begin{center}                         
\epsfig{file=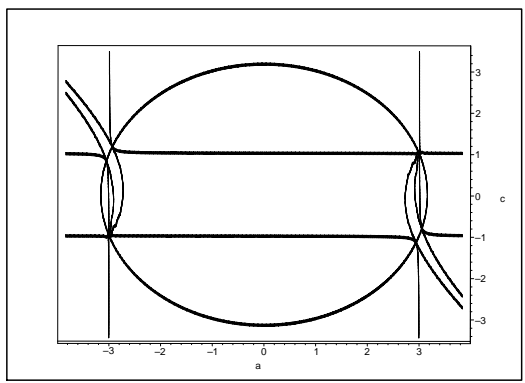,angle=270,width=0.5\textwidth}
\end{center}                         
\vspace{-2mm}\caption{A return to the triply connected ${\cal
D}^{(H)}$ at $b=0.1$.
 \label{fi6}}
\end{figure}

The change of the pattern is finally  achieved slightly below
$b=0.4$, at the moment when both of the new intersection candidates
touch the circle $\mathbf{C}(b) =
\partial {\cal D}^{(P)}|_{b=fixed}$ in a single point.
Subsequently, this point splits into the pair of the triple
intersections and the further decrease of $|b|$ forms the pattern
which is sampled in Fig.~\ref{fi7} at $b=0.2$ and in Fig.~\ref{fi6}
at $b=0.1$.

\begin{figure}[htb]                     
\begin{center}                         
\epsfig{file=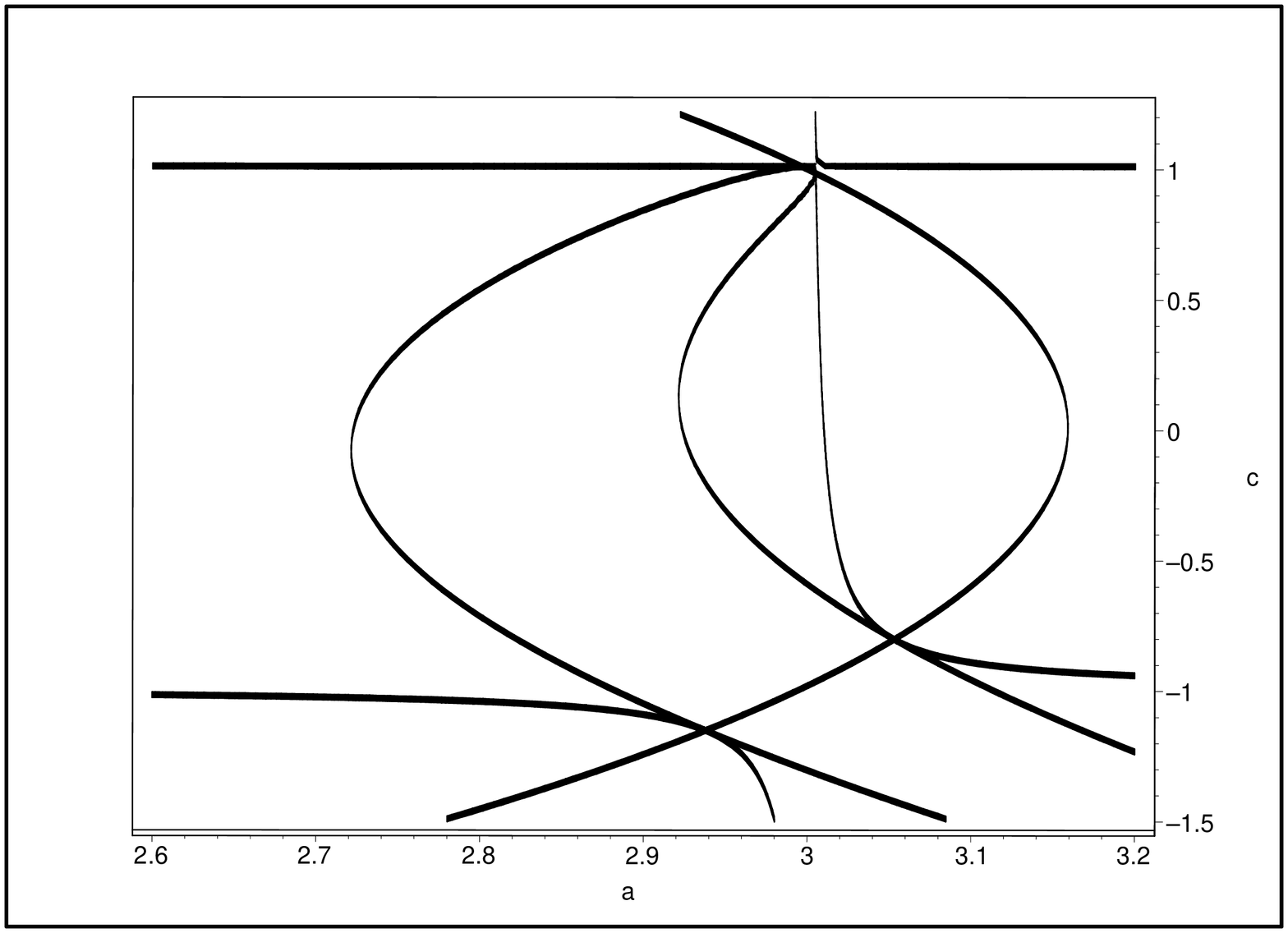,angle=270,width=0.5\textwidth}
\end{center}                         
\vspace{-2mm}\caption{A magnified detail of Fig.~\ref{fi6}.
 \label{fi6a}}
\end{figure}

\begin{figure}[htb]                     
\begin{center}                         
\epsfig{file=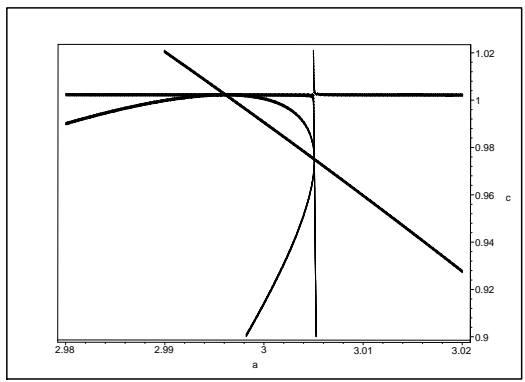,angle=270,width=0.5\textwidth}
\end{center}                         
\vspace{-2mm}\caption{A magnified detail of Fig.~\ref{fi6a}.
 \label{fi6aa}}
\end{figure}

Certainly, the key features of the new situation are better visible,
at the illustrative $b=0.1$, in its magnified presentation as
mediated by Figs.~\ref{fi6a} and~\ref{fi6aa}). Now we may return to
the limiting pattern of Fig.~\ref{fi5} where we witness the abrupt
change of the topology caused by the final confluence of the
straight and backbanding branch of the boundary $\partial {\cal
D}^{(W)}|_{b=fixed}$ in the limit $b\to 0$.

Obviously, in a way confirmed by the complementary results of
section \ref{juraj} such a confluence of the boundaries only occurs
in the limit so that the thee-dimensional version of the open set
${\cal D}^{(H)}$ is ultimately confirmed to be triply connected.

\section{Summary}

In the history of pure mathematics the specification of the horizons
$\partial {\cal D}^{(H)}$ (called, often, ``discriminant surfaces"
in this context) has been perceived as a challenging and rather
difficult problem even in its first ``unsolvable" case
characterized, in our present notation, by the Hilbert-space
dimensions $N = 5$ \cite{ams}. It is also certainly worth noticing
that in certain mathematically natural directions the real progress
is of an amazingly recent date \cite{amsdve}. In this connection it
seems rather remarkable that certain parallel developments occured
also in several applied-mathematics oriented studies paying
attention to the natural presence of more symmetries in the
Hamiltonian \cite{horizon} and/or to the introduction of more
observable quantities within a phenomenological quantum model in
question \cite{cc}.

In a constructive, more pragmatic setting as sampled by our recent
paper \cite{anom} we restricted our attention to the topological
aspects of the problem of horizons. The most obvious motivation of
such an effort has been given by the fact that a disconnectedness of
the domain ${\cal D}^{(H)}$ immediately requires the transition from
its traditional perturbation-theory descriptions (with a recommended
recent compact sample given in \cite{Langerplus}) to
non-perturbative methods, or to certain perturbative strong-coupling
techniques at least \cite{Rafa}. In Ref.~\cite{anom} the parallel
and less formal motivation has been emphasized to lie in the
systematic search for possible physical origin of the {\em
dynamical} anomalies in a {\em kinematical} nontriviality of the
topology of phase space.

The conclusions of our present paper are encouraging. Firstly we
demonstrated that for many purposes it may be sufficient to use the
matrices with a not too large $N$. Secondly, we have shown an
increase of the feasibility of the description of models $H$ with
certain additional symmetries. At the first nontrivial Hilbert-space
dimension $N=4$ they reduced, for example, the minimal necessary
number of parameters to $D=3$ or even to $D=2$ .

Thirdly we clarified that once one works with a tridiagonal-matrix
form of the ``unperturbed" $N$ by $N$ Hamiltonian $H^{(N)}_0$
complemented by certain computationally suitable {\em specific}
perturbations, the existence of the disconnected subdomains in
${\cal D}^{(H)}$ opens the direct access to the strong-coupling
dynamical regime.

Fourthly, on mathematical side, we are now able to recommend,
strongly, the use of certain auxiliary symmetries in the
Hamiltonians (e.g., the ones recommended in Ref.~\cite{maximal}). In
such cases, the algebraic secular equations pertaining to the model
often happen to factorize, leading to the much better tractable
polynomial equations of perceivably lower orders.

Last but not least it seems worth emphasizing that even at our
present ``solvable" choice of the dimension $N=4$ the latter fact
rendered our toy model more easily solvable. In fact, on the
background given by Refs. \cite{anom,ctyri} it took some time for us
to imagine that the anomalies of spectra could certainly start
occurring at the very small dimensions  $N<8$. In this context the
message of our present $N=4$ study is encouraging. Partly because
several specific spectral irregularities as observed at $N=8$ in
Ref.~\cite{anom} do also exist for certain very similar Hamiltonian
matrices with the dimension as low as $N=4$, and partly because our
model reconfirmed the hypothesis of a ``hidden" topology-related
connection between the loop-shaping of the lattices (i.e.,
presumably, Betti numbers in continuous limit) and the existence of
definite strong-coupling dynamical anomalies in the spectra of the
energy levels.

\section*{Appendix A. The three-Hilbert-space formulation of quantum mechanics}

In section 2 of Ref.~\cite{Wang} one finds one of the most compact
introductions into the abstract formalism of ${\cal PT}-$symmetric
quantum mechanics (PTSQM). Thus, we may shorten here the
introductory discussion and refrain ourselves to a few key comments
on the general theoretical framework.

In such a compression the PTSQM formalism may be characterized as
such a version of entirely standard quantum mechanics in which, in
principle, the system in question is defined in a certain
prohibitively complicated physical Hilbert space of states ${\cal
H}^{(P)}$ where the superscript may be read as abbreviating
``prohibited" as well as ``physical" \cite{SIGMA}.

Typical illustrative realistic examples may be sought in the physics
of heavy nuclei where the corresponding fermionic states are truly
extremely complicated. In the latter exemplification the first half
of the PTSQM recipe lies in the transition to a suitable, unitary
equivalent Hilbert space,  ${\cal H}^{(P)}\to {\cal H}^{(S)}$, where
the superscript ``$^{(S)}$" may stand for ``suitable" or ``simpler"
\cite{SIGMA}.

In the above-mentioned realistic-system illustration, for example,
the new space ${\cal H}^{(S)}$ coincided with a suitable
``interacting boson model" (IBM). In the warmly recommended review
paper of this field \cite{Geyer} it has been emphasized that the
requirement of the unitary equivalence between the two Hilbert
spaces ${\cal H}^{(P)}$ and ${\cal H}^{(S)}$ may only be achieved in
two ways. Either the corresponding boson-fermion-like mapping
$\Omega$ between these two Hilbert spaces (known, in this context,
as the Dyson's mapping) remains unitary (and the mathematical
simplification of the problem remains inessential) or is admitted to
be non-unitary (which is a less restrictive option which may enable
us to achieve a really significant simplification, say, of the
computational determination of the spectra).

What remains for us to perform and explain now is the second half of
the general PTSQM recipe. Its essence lies in the weakening of the
most common unitarity requirement imposed upon the Dyson's mapping,
 \ben
 \Omega^\dagger = \Omega^{-1}
 \een
to the mere quasi-unitarity requirement
 \ben
 \Omega^\dagger = \Theta\,\Omega^{-1}\,.
 \een
The symbol $\Theta\neq I$ represents here the so called metric
operator which defines the inner product in Hilbert space ${\cal
H}^{(S)}$.

More details using the present notation may be found in
\cite{SIGMA}. Just a few of them have to be recalled here. Firstly,
the main source of the purely technical simplifications of the
efficient numerical calculations  (say, of the spectra of energies)
is to be seen in the introduction of the third, purely auxiliary
Hilbert space ${\cal H}^{(F)}$ where the superscript ``$^{(F)}$"
combines the meaning of ``friendlier" with ``falsified"
\cite{SIGMA}.

By definition, the two Hilbert spaces ${\cal H}^{(S)}$ and ${\cal
H}^{(F)}$ coincide as the mathematical vector spaces (``of ket
vectors" in the Dirac's terminology). One only replaces the
nontrivial metric $\Theta^{(S)}\ \equiv\ \Omega^\dagger\Omega$ of
the former space by its trivial simplification  $\Theta^{(F)}\
\equiv\ I$ in the latter Hilbert space. As an immediate consequence,
the latter space acquires the status of an auxiliary, manifestly
unphysical space which does not carry any immediate physical
information or probabilistic interpretation of its trivial though,
at the same time, maximally mathematically friendly inner products.

\section*{Appendix B. The role of ${\cal PT}-$symmetry }

In its most widely accepted final form described in Ref.~\cite{Carl}
the PTSQM recipe complements the latter general scheme by another
assumption. It may be given the mathematical form of the
introduction of the second auxiliary, manifestly unphysical vector
space  ${\cal K}^{({\cal P})}$ which is, by definition, not even the
Hilbert space. In fact, this fourth vectors space is assumed endowed
with the formal structure of the so called Krein space
\cite{Azizov}.

Ref.~\cite{Langerplus} may be consulted for more details. Here, let
us only remind the readers that the symbol ${\cal P}$ in the
superscript carries the double meaning and combines the mathematical
role of the indefinite metric ${\cal P}$ (defining in fact the Krein
space) with an input physical interpretation (usually, of the
operator of parity). In addition, the theoretical pattern
 \be
 {\cal H}^{(S)} \leftrightarrow {\cal K}^{({\cal P})}
 \leftrightarrow {\cal H}^{(F)}\,.
 \label{pattern}
 \ee
gets complemented by the requirement that there exists a ``charge"
operator ${\cal C}$ such that the (by assumption, non-trivial,
sophisticated) metric $\Theta^{(S)}\neq I$ which defines the inner
product in the second Hilbert space ${\cal H}^{(S)}$ coincides with
the product of the two above-mentioned operators,
 \be
 \Theta^{(S)}={\cal PC}\,.
 \ee
The contrast between the feasibility of the $N \neq 3$ constructions
presented in Ref.~\cite{Wang} and the discouraging complexity and
incompleteness of the next-step $N=4$ constructions as performed in
our older paper \cite{ctyri} and in its sequels \cite{horizon} was
also thoroughly discussed in our review \cite{acta}. In our present
text we do not deviate from the notation and conventions accepted of
this review. We solely pay attention to the class of models where
$N-$dimensional matrix of parity is unique and given, in advance, in
the following form,
 \be
 {\cal P}=
 \left[ \begin {array}{cccccc}
   1&0&\ldots&&\ldots&0  \\
  0 &-1&0&\ldots&\ldots&0 \\
 0&0&1&0&\ldots&0 \\
 0&0&0&-1&\ddots&\vdots \\
  \vdots&\ddots &\ddots&\ddots&\ddots&0 \\
  0&\ldots&0&0&0&\mp 1
   \end {array} \right]\,.
   \label{parita}
 \ee
In parallel, we did make use of the time-reversal operator ${\cal
T}$ of the form presented, e.g., in Ref.~\cite{Wang} as mediating
just the transposition plus complex conjugation of vectors and/or
matrices. One should add that once we  solely work with real vectors
and matrices, we are even allowed to perceive ${\cal T}$ as a mere
transposition.

\subsection*{Acknowledgments}

Work supported by the GA\v{C}R grant Nr. P203/11/1433, by the
M\v{S}MT ``Doppler Institute" project Nr. LC06002 and by the Inst.
Res. Plan AV0Z10480505.

\end{document}